\documentclass[a4paper,11pt]{article}
\pdfoutput=1

\usepackage{contribution}
\usepackage{isotope}
\usepackage{graphics}
\usepackage{graphicx}
\usepackage{epsfig}

\newcommand{\weblink}[2][]{%
    \ifthenelse{\equal{#1}{}}%
    {\textnormal{\url{#2}}}%
    {\textnormal{\href{#2}{#1}}}%
}

\def\beq{\begin{equation}}
\def\eeq#1{\label{#1}\end{equation}}
\def\eeqn{\end{equation}}

\def\beqa{\begin{eqnarray}}
\def\eeqa#1{\label{#1}\end{eqnarray}}
\def\eeqan{\end{eqnarray}}

\let\bar=\overbar

\def\Dslash{\not{\hbox{\kern-4pt $D$}}}
\def\dslash{\not{\hbox{\kern-2pt $\del$}}}

\def\msb{{\bar{\ssstyle M \kern -1pt S}}}

\newcommand{\contribution}[7][]{%
  \clearpage
  \thispagestyle{plain}
  \ifthenelse{\equal{#1}{}}
  {\hypersetup{pdftitle={#2}}}
  {\hypersetup{pdftitle={#1}}}
  \hypersetup{pdfauthor={{#3} {#4}}}
  {\centering\normalfont\LARGE\bfseries\sffamily #2 \par\nobreak}
  \lhead{}
  \chead{%
    \textit{\footnotesize XIV International Conference on Hadron Spectroscopy
      (\weblink[\textit{hadron2011}]{http://www.hadron2011.de}), 13-17 June 2011, Munich, Germany}%
  }
  \rhead{}
  \bigskip
  \begin{center}
    {#3} {#4}\ifthenelse{\equal{#6}{}}{}{\footnote{\weblink[#6]{mailto:#6}}}
    \ifthenelse{\equal{#7}{}}{}{#7} \\
    \textit{#5}
  \end{center}
  \bigskip
}

\renewcommand{\abstract}[1]{%
  \begin{center}
    \begin{minipage}{0.85\textwidth}
      \begin{footnotesize}
        #1
      \end{footnotesize}
    \end{minipage}
  \end{center}
  \bigskip
}

\begin{document}

%
%
%
%
%
{  


%

\contribution[Exotic Mesons in $\gamma p\to\pi^+\pi^+\pi^-n$]  
{The Search for Exotic Mesons in $\gamma p\to\pi^+\pi^+\pi^-n$ with CLAS at Jefferson Lab}  
{Craig}{Bookwalter}  
{Department of Physics \\
 Florida State University \\
 Tallahassee, FL 32306}  
{craigb@jlab.org}  
{on behalf of the CLAS Collaboration}  
%

\abstract{%
\par The $\pi_1(1600)$, a $J^{PC} = 1^{-+}$ exotic meson has been observed by experiments using pion beams.
Theorists predict that photon beams could produce gluonic hybrid mesons, of which the $\pi_1(1600)$ is a candidate, at enhanced levels relative to pion beams. 
The g12 rungroup at Jefferson Lab's CEBAF Large Acceptance Spectrometer (CLAS) has recently acquired a large photoproduction dataset, using a liquid hydrogen target and tagged photons from a 5.71 GeV electron beam.  
A partial-wave analysis of 502K $\gamma p \to \pi^+\pi^+\pi^-n$ events selected from the g12 dataset has been performed, and preliminary fit results show strong evidence for well-known states such as the $a_1(1260)$, $a_2(1320)$, and $\pi_2(1670)$.
However, we observe no evidence for the production of the $\pi_1(1600)$ in either the partial-wave intensities or the relative complex phase between the $1^{-+}$ and the $2^{-+}$ (corresponding to the $\pi_2$) partial waves.
}
%


\section{Introduction}
\par Theoretical work indicates that photon beams may be able to abundantly produce gluonic hybrid mesons.
The flux-tube model of Isgur and Paton \cite{Isgur:1983wj} has been central in the theoretical study of gluonic hybrids,
and the vector nature of the photon is optimal for promoting the flux-tube to an excited state, possibly producing hybrids in proportions equal to that of the $a_2(1320)$ \cite{Close:1994pr}. 
Additionally, calculations on the lattice in the charmonium regime show a strong photocoupling for $c\bar{c}$ hybrids\cite{Dudek:2009kk}, which could bode well for the photoproduction of light exotics.
Some of these states carry spin $J$, parity $P$, and $C$-parity that are inaccessible to ordinary $q\bar{q}$ matter.
These exotic $J^{PC}$ states are readily identified in experiment by means of partial-wave analysis.
Light exotics have been searched for in pion production but photoproduction has remained largely unexplored until recently.
\par Members of the CLAS collaboration in Hall B at Jefferson Lab have completed two photoproduction experiments to look for light exotics.
The first, completed in 2001 as part of the CLAS g6c rungroup, acquired 250K $\gamma p\to\pi^+\pi^+\pi^-n$ events in search of the $\pi_1(1600)$ exotic, sighted first in pion production experiments \cite{Chung:2002pu}, \cite{Alekseev:2009xt}.  
CLAS acceptance is optimized for baryon spectroscopy, so the partial-wave analysis was performed on a smaller set of 83K events where background from $\Delta$ and $N^*$ decays had been removed by kinematic cuts.
No evidence for a $1^{-+}$ exotic signal was found \cite{:2008bea}, a direct challenge to claims that gluonic hybrids could be produced on an equal footing with the $a_2(1320)$.
However, statistics were insufficient to rule out a $\pi_1(1600)$ produced at the same level as it was observed in pion production, at a few percent of the $a_2$. 
\par In order to look for $\pi_1(1600)$ photoproduction at lower levels, the HyCLAS experiment was proposed in 2004, with data taken in 2008 as a member of the CLAS g12 rungroup. 

\section{Features of the CLAS g12 Dataset}

\par The g12 run of CLAS was completed in June of 2008, acquiring 26 billion events of various topologies. 
From these 26B triggers we have isolated 6 million exclusive $\gamma p\to \pi^+\pi^+\pi^- n$ events by vertex and timing cuts, with the neutron selected via missing mass.
These events have either a mesonic topology, where $\gamma\;p\to\;Xn\to\pi^+\pi^+\pi^-n$, or a baryonic topology, where $\gamma\;p\to\;N^*\rho\to\pi^+\pi^+\pi^-n$, as can be seen by the light-shaded distributions in Figure \ref{fig:m_npi}.
Our analysis depends on having a pure sample of mesonic events, so we selected events with the following criteria:
\begin{center}
  \begin{tabular}{c c}
    $|t_{3\pi} - t_{min}| = |t'| < 0.105$ GeV$^2/c^4$; & $\theta_{lab}(\pi^+) < 25^\circ$ (both $\pi^+$) \\
  \end{tabular}
\end{center}
An example of the post-cut $n\pi$ mass distributions, as well as the resulting 3$\pi$ mass distribution can also be found in Figure \ref{fig:m_npi}.

\begin{figure}
  \begin{center}
    \includegraphics[width=6in]{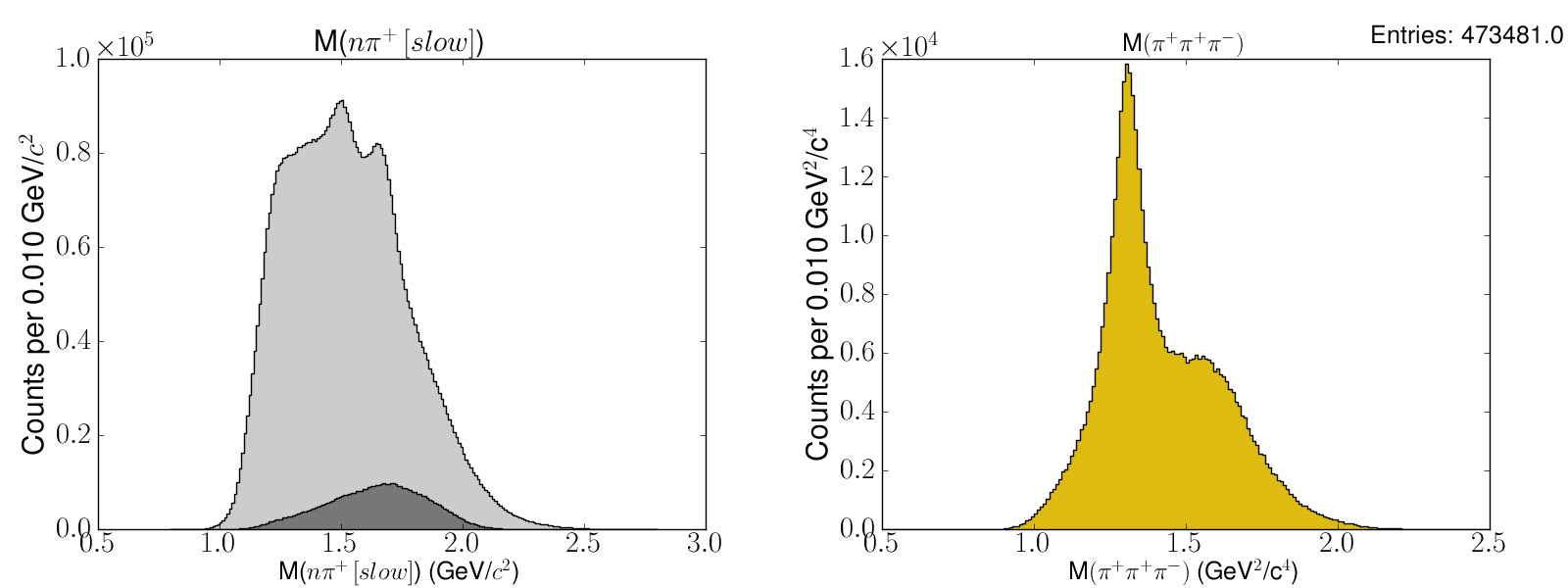}
    \caption{\textbf{Left}, the invariant mass of the neutron-$\pi^+$ system for the slow $\pi^+$, where the lighter pre-cut region shows $N^*$ peaks and the darker region is what remains after kinematic cuts.
      \textbf{Right}, the 3$\pi$ invariant mass distribution after the kinematic selections are applied. The prominent peak is in the region where one would expect an $a_2(1320)$, while a shoulder exists in the $\pi_2(1670)$ region.
      \label{fig:m_npi}}
  \end{center}
\end{figure}


\section{Partial-Wave Analysis of $\gamma\;p\to\pi^+\pi^+\pi^-n$ in CLAS g12 Data} 

\begin{figure}
\begin{center}
\includegraphics[width=5in]{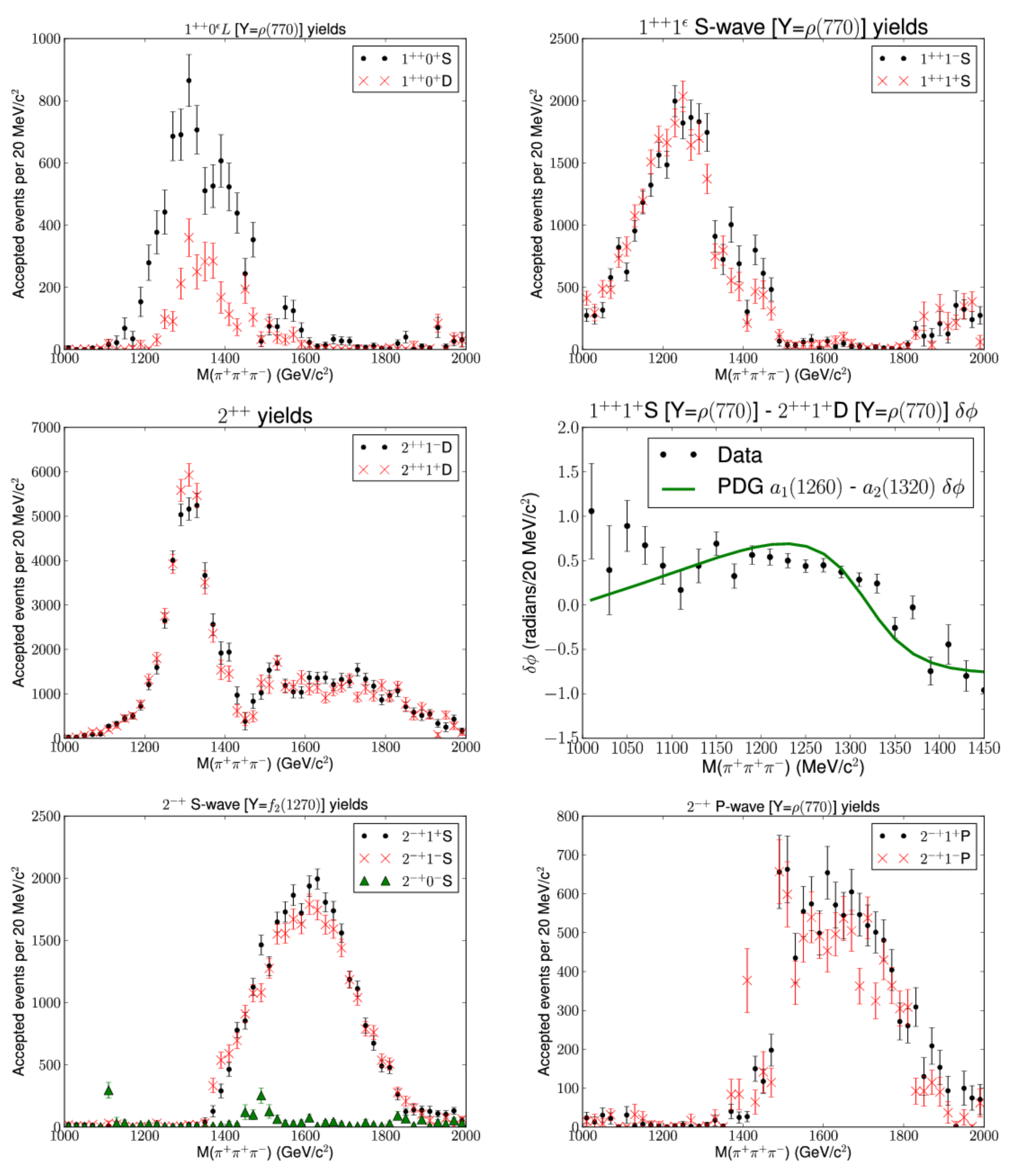}
\caption{\textbf{Top row}: accepted yields for $1^{++}$ waves. The yields for the M=0 waves on the left are a possible signature for the Deck effect in photoproduction, as a photon beam exchanging a $\pi$ forbids any states from being produced with M=0. On the right, the $(\rho\pi)_S$ decay of the $1^{++}$ shows strong evidence for the $a_1(1260)$. 
\textbf{Middle row}: on the left, the accepted yields of the $2^{++}$, showing strong evidence for the presence of the $a_2(1320)$. The two reflectivities are populated equally because of the circular polarization of the photon beam used in the experiment. On the right, the phase difference between the $1^{++}1^+$S and $2^{++}1^+$D waves. Overlaid is the phase difference between two Breit-Wigner amplitudes constructed with the mass and width reported by the Particle Data Group. 
\textbf{Bottom row}: accepted yields for the largest $2^{-+}$ contributors. The intensity for the $(f_2\pi)_S$ decay on the left, along with the $(\rho\pi)_P$ decay on the right, shows good evidence for the presence of the $\pi_2(1670)$.
\label{fig:nonexotic}}
\end{center}
\end{figure}

\par We then subjected the final sample of 502K events to a partial-wave analysis based on the helicity formalism.
Decay amplitudes are calculated in the reflectivity basis to ensure parity conservation, and then each is paired with a complex production amplitude.
Those production amplitudes were then varied in a likelihood fit to find the most probable mixture of states given the set of input events.
We then examine the norm of each production amplitude to look for peaks, as well as examining the phase difference between pairs of production amplitudes to look for movement corresponding to the interference of two Breit-Wigner distributions.
\par In particular, the fit presented contains 19 waves spread among four $J^{PC}$: $1^{++}$, $2^{++}$, $1^{-+}$, and $2^{-+}$.
Since the helicity of the photon is constrained to be $\pm 1$ and $\pi$ exchange is the dominant production mechanism, we should not observe any spin-0 states, one of the most complicated features of pion-production analyses.
Thus, most of the intensity is allocated to the M=1 waves.   
However, in the $1^{++}$ waves, shown in the first row of Figure \ref{fig:nonexotic}, the M=0 makes a significant contribution. 
It is possible that these events are due to a $\rho\pi$ system produced in the S-wave via a Deck process.
The M=1 waves of the $1^{++}$ show evidence for the $a_1(1260)$ in the S-wave $\rho\pi$ decay; interpretation of the $1^{++}$D M=1 waves is not as clear presently.

\par The $2^{++}$ waves, illustrated in the middle row of Figure \ref{fig:nonexotic}, show strong evidence for the presence of the $a_2(1320)$ in this data sample. 
The $2^{++}$D M=0 wave was omitted because when included, it has zero intensity, as one would expect for an $a_2(1320)$ produced via pion exchange from a photon beam.
We can test our resonant interpretation of the $1^{++}$S and $2^{++}$D intensities by examining their phase difference. The results, illustrated in the middle right plot of Figure \ref{fig:nonexotic}, seem to indicate the presence of the $a_1(1260)$ and $a_2(1320)$ in our sample.
\par The $2^{-+}$ waves, illustrated in the bottom row of Figure \ref{fig:nonexotic}, show strong evidence for the presence of the $\pi_2(1670)$ in our sample as well. 
Both the $2^{-+}$ $(f_2(1270)\pi)_S$ and $2^{-+}$ $(\rho\pi)_P$ show broad enhancements reminscent of Breit-Wigner shapes, with the $f_2\pi$ decay the dominant mode, as one would expect from branching ratios reported by the PDG.
Also if the Deck effect is present in this channel, we should also observe it in the $2^{-+}$ M=0 waves. 
Only the $(f_2\pi)_S$ M=0 wave is included in this fit, but one can indeed see a small peak in its intensity.
\begin{figure}
\begin{center}
\includegraphics[width=2.5in]{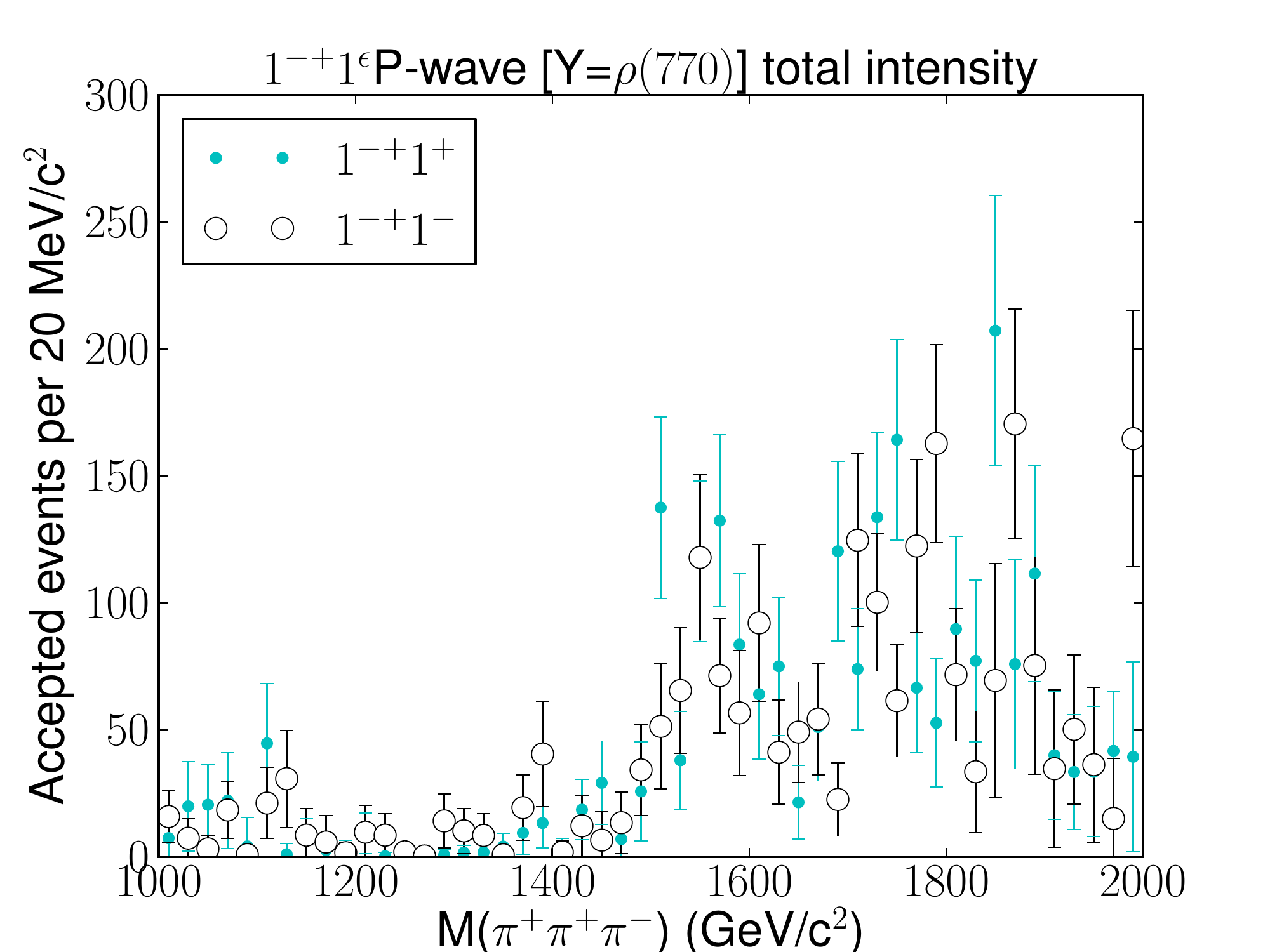}
\includegraphics[width=2.5in]{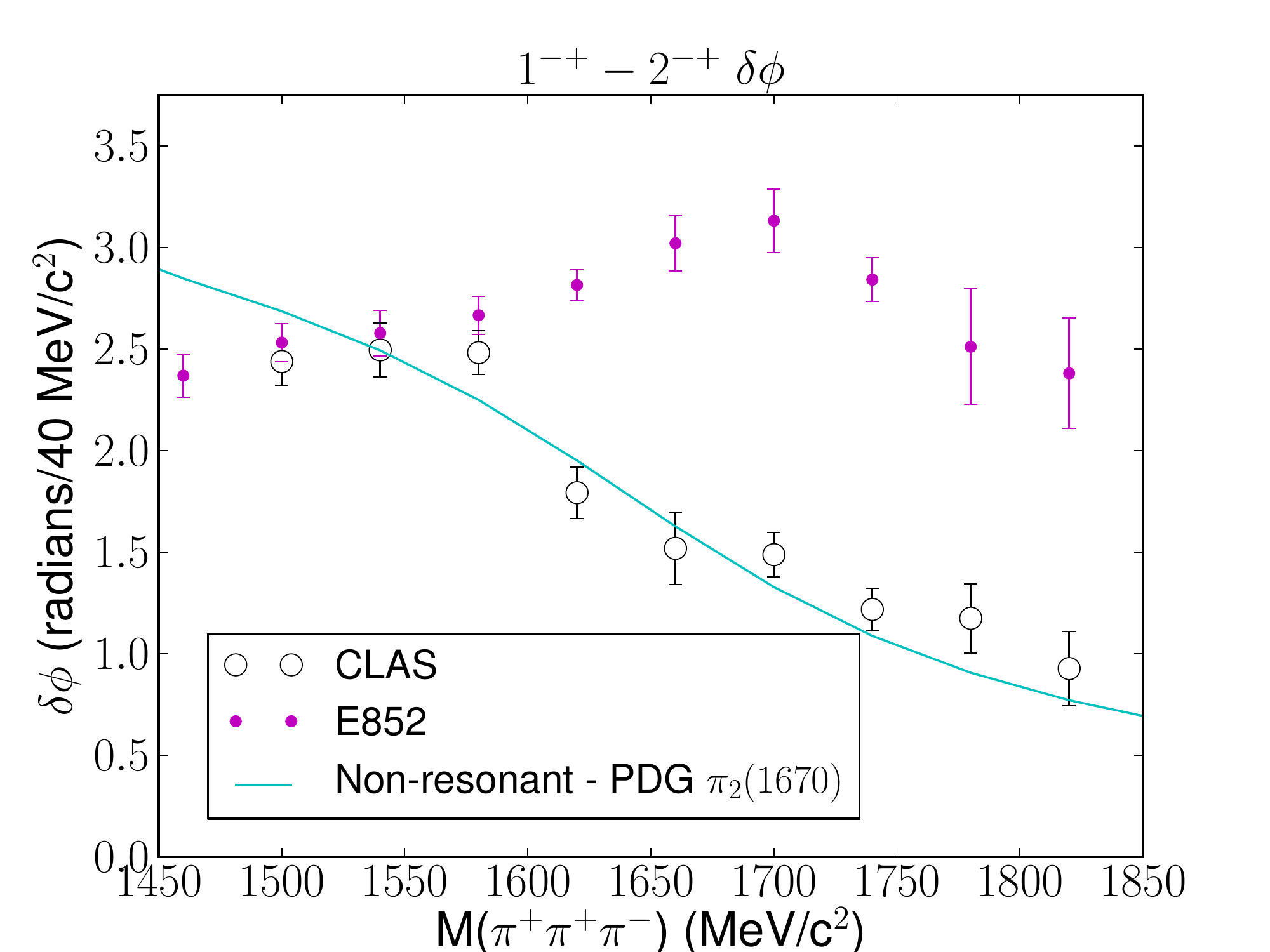}
\caption{On the \textbf{left}, the intensities of the $1^{-+}$ M=1 waves are plotted, showing no clear evidence for resonant structure, as is true for the $1^{-+}$ M=0 wave (not shown). On the \textbf{right} is an overlay of the $1^{-+}$ - $2^{-+}$ phase motion for both the presented CLAS result and the results reported by E852 in \cite{Chung:2002pu}. One can see a clear turning-over of the E852 phase, indicative of interference between two Breit-Wigner forms. The CLAS phase has a steady decrease, indicative of a resonating $\pi_2$ being subtracted from a nonresonant $1^{-+}$, as shown by the curve.
\label{fig:pi1}}
\end{center}
\end{figure}

\par Finally, the exotic $1^{-+}$ waves, as illustrated in Figure \ref{fig:pi1}, do not show resonant structure in their intensities.
Furthermore, examining the phase difference relative to the $2^{-+}$ M=1 $(f_2\pi)_S$ waves, one finds no motion indicative of a resonance.
In fact, as also illustrated in Figure \ref{fig:pi1}, one can compare directly between the phase observed in \cite{Chung:2002pu} and the phase we observe, and where there is a clear turning-over in the E852 data, our data shows a clear downward trend, indicative of a resonant $2^{-+}$ subtracted from a nonresonant background. 
\par Thus our preliminary conclusion is that there is no evidence for the presence of a $1^{-+}$ resonance in our data sample. 
These results are not necessarily in conflict with past pion-production results; the analyses in \cite{Chung:2002pu} and \cite{Alekseev:2009xt} examine diffractive processes while this analysis proceeds via charge exchange.
Thus we can explain the discrepancy if we posit that the $\pi_1(1600)$ is produced via Pomeron exchange.



%

}  


\end{document}